\documentstyle[11pt,newpasp,twoside,epsf]{article}
\def\edcomment#1{\iffalse\marginpar{\raggedright\sl#1\/}\else\relax\fi}
\reversemarginpar

\begin{document}
\title{Using {\em Chandra}/LETG spectra to probe stellar coronae}
 \author{Gaitee A.J. Hussain}
\affil{Astrophysics Division, Research and Science Support Department of ESA, ESTEC, Postbus 299, NL-2200 AG Noordwijk, The Netherlands}
\author{Nancy Brickhouse, Andrea K. Dupree, Adriaan A. van Ballegooijen}
\affil{Harvard Smithsonian CfA, 60 Garden St, Cambridge MA02138, USA}
\author{Andrew Collier Cameron, Moira Jardine}
\affil{School of Physics \& Astronomy, Univ. of St Andrews, Fife, UK}
\author{Jean-Francois Donati}
\affil{Laboratoire d'Astrophysique, Observ. Midi-Pyr\'en\'ees,  Toulouse, France}

\begin{abstract}

We probe the relationship between surface magnetic fields and 
the X-ray emitting corona in the rapidly rotating star AB Dor.
Circularly polarised spectra have been inverted to produce a 
                  surface (photospheric) magnetic field map. 
This surface map has been
extrapolated to model AB Dor's
coronal field topology and  X-ray light curve. 
{\em Chandra}/LETG light curves of AB Dor from the same epoch
 show intrinsic variability at the 30\% level.
Period analysis indicates a fraction of this is due  
to rotational modulation. We measure  velocity shifts in 
emission line centroids 
as a function of ${ P}_{\rm rot}$ and
find evidence of rotational modulation 
(max.~vel. $\sim 40 \pm 13$\,km\,s$^{-1}$). This modulation may 
indicate the presence of a  localised X-ray emitting region at mid-to-high 
latitudes.
\end{abstract}

\section{Introduction}
We present the first results from a co-ordinated optical 
and X-ray campaign to probe the corona of a rapidly 
rotating, active cool star. New 
X-ray and EUV observations 
reveal a picture of stellar coronae unlike anything on the Sun.
Emission measures of about $10^{52}$\,cm$^{-3}$ 
 are found in plasma at temperatures of
8\,MK and beyond, even in relatively quiescent coronae
(e.g. Sanz-Forcada et al. 2003).
Emission measures peak at $10^{50}$\,cm$^{-3}$ at 2.5\,MK
on the Sun  {\em in active regions} 
(Drake et al. 2000).
While the thermal properties of  active stellar
 coronae are increasingly well-determined,
we have yet to establish where this emission originates and to
confront observations with theoretical models.

Previous studies provide evidence for both extended and 
compact coronae in active cool stars.
The extended coronal model is supported by 
observations of fast moving cool ($10^4$\,K)
H$\alpha$ absorption transients in young stars.
These are caused by prominences rigidly corotating with the star at heights 
beyond the Keplerian corotation radius
(Collier Cameron et al. 2001). 
Further support for extended coronae
comes from time-resolved HST spectroscopy of the eclipsing binary, 
V471 Tau (K2V+DA). The hot white dwarf acts as
a probe of circumstellar material around the K2V 
enabling Walter ({\em this volume})
to detect transition region temperature plasma  
($10^4<T<10^5$\,K) above  2\,$R_*$.

\begin{figure}
\plotone{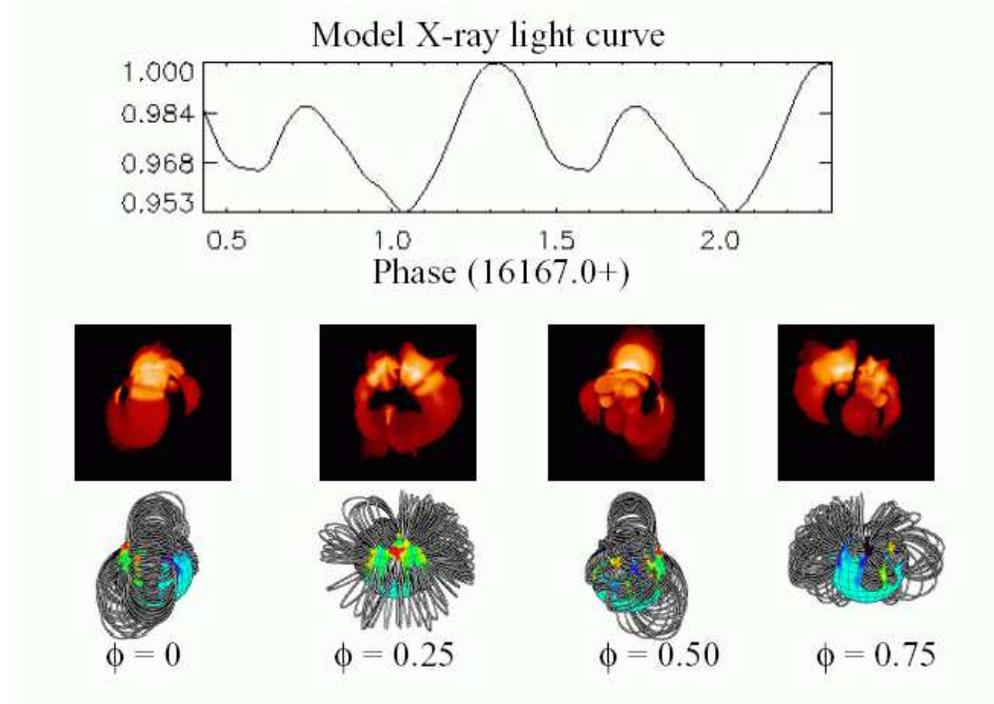}
\caption{[{\em bottom row}] Surface magnetic field maps  from 2002~Dec~11 
are extrapolated to model AB Dor's 3-D coronal field,  generate
``X-ray snapshots''   [{\em middle row}], and  predict its 
X-ray light curve  [{\em top row}].
}
\end{figure}

Support for compact coronae comes from 
X-ray light curves of the eclipsing binary, 
Algol (B8V+K2III). Flaring  
regions are found to have heights, 
$0.1<H<0.6$\,$R_*$ 
above the K star (Schmitt \& Favata 1999, Schmitt et al. 2003).
Maggio et al. (2000) find 
evidence for a small flaring region ($H<0.3$\,$R_*$)
on the single, K0V, AB Dor.
likely to be located near its pole.
Electron density diagnostics measured in $10^7$\,K plasma
in active stars are typically very high ($n_e>10^{12}$\,cm$^{-3}$).
Densities this high are more easily explained in terms of 
 compact X-ray emitting regions 
  (e.g. Sanz-Forcada et al. 2003).

Brickhouse et al. (2001) find evidence for 
compact, quiescent coronae in active stars in a 
{\em Chandra}/HETG study of the contact binary, 
44i~Boo (G0V~\&~G0V), spanning over 2.5\,${P}_{\rm rot}$.
Flaring emission is filtered out and velocity shifts in  line
centroids of emission line profiles are measured as a function of 
rotation phase.  
Rotational modulation in the quiescent corona is detected, 
indicating  a compact emitting area  near the pole of the primary. 
We  want to follow up on this latter
study in order to learn whether  the 
picture of compact stellar coronae is applicable to other active cool stars 
or whether it is exclusive to contact binaries.

\subsection{AB Doradus} 
 
AB Doradus 
(K0V, $m_V$=6.9, ${P}_{\rm rot}$=0.51\,d, v$_e \sin i$=90\,km\,s$^{-1}$) 
is an active, rapidly rotating single star that has recently arrived 
onto the main sequence. It is an ideal candidate for 
spectroscopic mapping techniques such as Doppler imaging as 
its 12.4\,hour rotation period means that a 
large fraction of its surface can be tracked in one night, and its
rotationally broadened spectral line profiles enable detailed
surface maps to be reconstructed ($\sim 3$\deg\ longitude
resolution at the equator). 
Spot maps of AB Dor's surface typically show 
a dark spot covering its pole (polar cap) with smaller spots co-existing
at lower latitudes  (e.g. Donati et al. 1999). AB Dor is also bright in
 X-ray wavelengths, $L_{x}/L_{bol}~10^{-3}$,  and shows strong signs of coronal 
variability.
ROSAT X-ray light curves of AB Dor indicate 
rotational modulation at the 5-13\% level (Kuerster et al. 1997).

\section{Coronal field modelling}
A time-series of high-resolution circularly polarised spectra was acquired 
using the AAT/UCLES/Semel polarimeter setup between 2002 Dec 11 to 14.
These spectra are used to generate an X-ray model of AB Dor and
to predict its {\em quiescent} X-ray lightcurve at this epoch using the 
following steps  (also see Fig.~1).

First, circularly polarised spectra are 
inverted using Zeeman Doppler imaging
(Semel 1989; Donati et al 2000). 
The resulting magnetic field maps are extrapolated assuming
that the field is potential and  open (i.e. not emitting) beyond 3.4\,$R_*$ 
(Jardine et al. 2001). 
If the closed corona is isothermal and in hydrostatic equilibrium
the 3-D X-ray emission model of AB Dor is as depicted in Fig.~1.
X-ray ``snapshots'' like these 
are used to predict the quiescent X-ray light curve
of AB Dor between 2002 Dec 11 to Dec 14.
This model light curve predicts X-ray rotational modulation at the 
5\% level in AB Dor's quiescent corona. 

\section{{\em Chandra}/LETG spectra}
We observed AB Dor using the {\em Chandra}/LETG instrument 
on 2002 December 11 over  88~ksec (1.98\,$P_{\rm rot}$).
The LETG spectra span a wavelength range from $6<\lambda<174$\,\AA, 
with spectral resolution ranging from $460<R<2000$ 
($R=2000$ at the longest wavelengths). 
Measuring line centroids  as precisely
as possible is crucial for this analysis, so 
it would make sense to use the best spectrally resolved 
lines in this dataset.
In this regard, however, LETG spectra  present
two challenges. Firstly, the highest wavelength lines in the dataset are also
the noisiest with the strongest background  (Fig.~2b).
This clearly affects our centroiding capability.
The second challenge is that the LETG wavelengths 
are poorly calibrated (S.M. Chung {\em priv. comm.}). 
We compensate for these deficiencies in the following ways:

We measure the discrepancy, $\Delta \lambda$, 
between the observed wavelength position
of the strongest emission lines and 
the theoretical wavelength,  $\lambda_0$ 
(using the {\sc atomdb} database). 
This $\Delta \lambda$ 
is used to correct the wavelength position of 
the observed line profile 
and  $\lambda_0$ is used to convert the wavelength scale into velocity-space. 
The 88~ksec exposure is then divided up into eight quarter-phase bins. 
To improve our centroiding capability, 
we integrate over the six strongest lines in the high spectral
resolution region ($>88$\AA).
This integration is done by converting 
the six strongest lines between $88<\lambda<133$\AA\  
into velocity-space,  rebinning them 
 onto the same velocity scale and  then integrating the counts 
from all the lines (Brickhouse, {\em this volume}).  We 
measure line centroids by fitting Gaussians to 
the ``summed'' line at each of the eight quarter-phase bins.
The centroid of this summed line profile is measured with a precision of
$\pm 13$\,km\,s$^{-1}$ (Fig. 3).
In practise, we find that the strong O\,{\sc viii}\,18.97\,\AA\ line 
can be measured with an equivalent level of precision ($\pm 18$\,km\,s$^{-1}$)
despite its relatively poor resolution.

\begin{figure}
\plotone{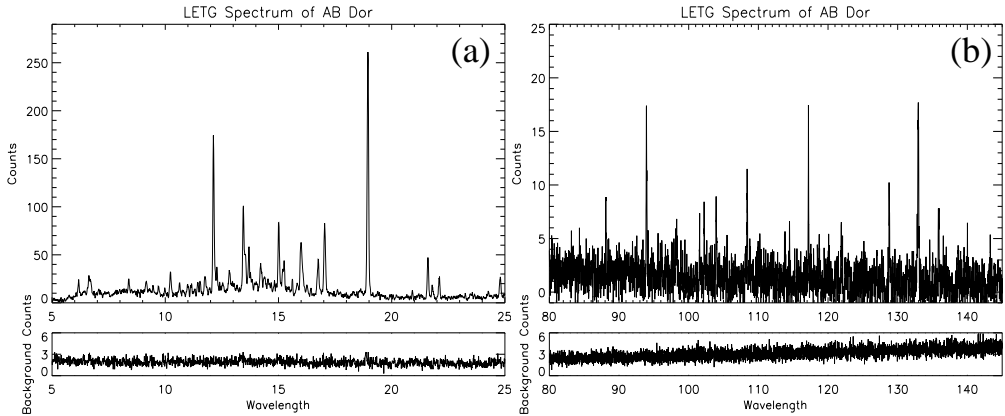}
\caption{{\em Chandra}/LETG spectrum integrated over 88\,ks. 
The upper panels, a\,\&\,b, show the low and high 
spectral resolution wavelength regions respectively.
The lower panels show the background levels.
}
\end{figure}

\begin{figure}
\plotone{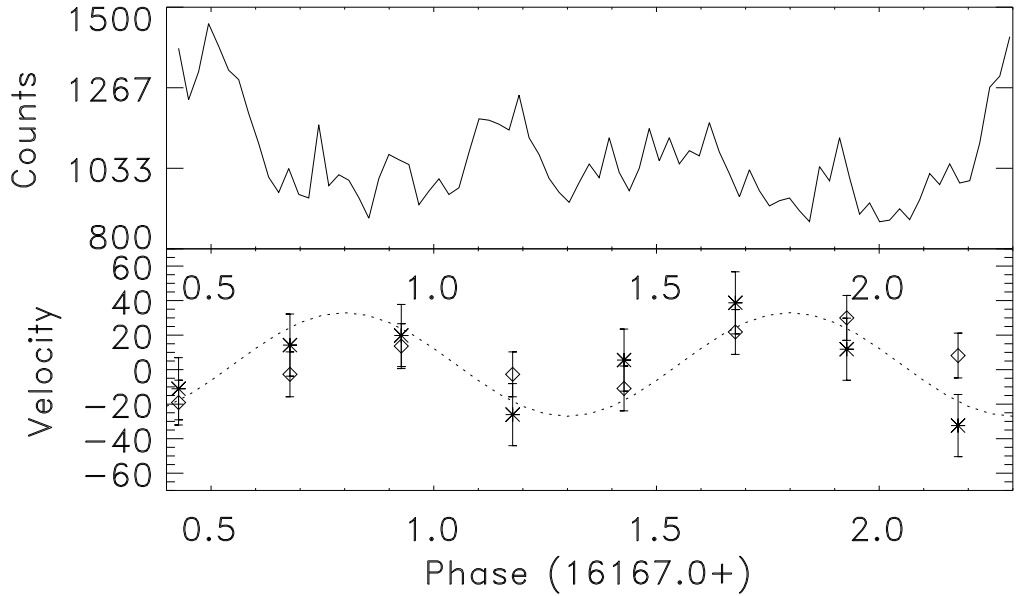}
\caption{[{\em Top panel}] The X-ray 
light curve varies by 30\%. 
[{\em Bottom panel}]
Velocity shifts in line centroids 
of the summed lines [$\circ$] and the 18.97\AA\ line [*]
indicate rotational modulation.}
\end{figure}

\section{Discussion}
Fig.~3 shows the X-ray light curve for AB Dor. We find that the light curve 
varies by 30\%. Initial periodogram analyses indicate that a fraction of
this is indeed due to rotational modulation. We will use
the method used by K\"urster et al. (1997) to filter   
this X-ray light curve in order 
to distinguish the rotationally modulated component. 
This rotationally modulated component
of the light curve will be compared with the predicted light curve in
Fig.~1. Our 3-D coronal model of AB Dor can also be refined, 
we can compensate for missing magnetic field information 
(e.g. in the polar cap). However, only a unipolar spot 
should significantly modify the coronal field distribution 
(McIvor et al. 2003). 
We will also evaluate the effect on the predicted X-ray lightcurve
if the global magnetic field is non-potential using the technique described
by Hussain et al. (2002).

Measurements of the centroids of the
best resolved spectral lines (the summed profile derived using the above
 method and  O\,{\sc viii}\,18\,\AA\ line) show evidence of 
rotational modulation that repeats from one
rotation cycle to the next. 
Even though both the O\,{\sc viii}\,18\,\AA\ and the summed line profile
show the same trend, 
the uncertainty associated with our velocity shift measurements  
 makes this result somewhat inconclusive.
This pattern of rotational modulation would indicate that 
the bulk of quiescent X-ray emission in AB Dor (at this epoch) 
originated in localised  regions, most likely at high latitudes 
($\sim$60$^{\circ}$).
Emission originating at the  pole would show no rotational modulation.
We will add more lines from the longer wavelengths  ($>88$\,\AA)
to reduce the error bars on this measurement.

\end{document}